\documentclass[12pt,a4paper]{article}

\usepackage[font=scriptsize]{caption}
\usepackage{titlesec}
\usepackage{multicol}
\usepackage{flushend}
\usepackage{amsfonts}
\usepackage{setspace}
\usepackage{balance}
\usepackage{nomencl}
\usepackage[numbers]{natbib}
\usepackage{amsmath}
\usepackage{amsthm}
\usepackage{amsmath}
\usepackage{amssymb}
\usepackage{float}
\usepackage{stmaryrd}
\usepackage{stackrel}
\usepackage{url}
\usepackage{color,hyperref}
\usepackage{subfig}
\usepackage{graphicx}
\usepackage{geometry}
\begin{document}



\title{Geodesics Structure and Light Deflection of Regular Phantom Black Hole}

\maketitle


\begin{center}
\author{{B. Malekolkalami} and {M. Haditale}}

\thanks{Faculty of Science, University of Kurdistan, Sanandaj, P. O. Box 416, Iran (email: B.Malakolkalami@uok.ac.ir, Maryam.haditale@uok.ac.ir)}
\end{center}

\begin{abstract}
The geodesic structure of  Regular Phantom Black Holes (\textbf{RPBH}) space--time is analyzed and discussed in three asymptotically cases: Flat, de Sitter (\textbf{dS}), and Anti--de Sitter (\textbf{AdS}). The impact of the important scale parameter $b$\footnote{This parameter determines the coupling  strength between phantom field and gravity.} on the trajectory of particles is studied and investigated which can mimic repulsion or attraction. By virtue of Effective Potential (\textbf{EP}) tool, the circular orbits and their stability are discussed. Also, in the  asymptotically flat spacetime, the angle of light deflection versus the scale parameter is presented.
\end{abstract}
\section{Introduction}\label{sec-1}
Cosmological  observations (over the last few decades) indicate that the universe is expanding at an accelerating rate,  as evidenced by observations of type Ia supernovae \cite{Airy1}, Cosmic Microwave Background (\textbf{CMB}) measurements \cite{Airy2, Airy3, Airy4}, and studies of the large-scale structure \cite{Airy5}. This accelerated expansion has raised concerns within the scientific community. While gravity typically acts to slow down the expansion rate, the observed accelerating expansion suggests the presence of an unknown substance known as Dark Energy (\textbf{DE}) in the universe \cite{Airy6}. Various DE models have been proposed, with the cosmological constant being one of the most well-known. However, there remain two unresolved issues with this model: the challenge of extracting vacuum energy from quantum field theory and the discrepancy in the amounts of DE and Dark Matter (\textbf{DM}) present in the universe \cite{Airy7}. Several astrophysical observations have emphasized the pressure to density ratio, such as a model-free data analysis of 172 type Ia supernovae (SNIa) yielding a diverse range of values \cite{Airy8}. The Planck data over specific years also provided insights \cite{Airy9}. An analysis of hot gas in 26 X-ray luminous and dynamically relaxed galaxy clusters using data from the Chandra Telescope was conducted \cite{Airy10}. Estimates on SNIa data from the SNLS3 sample were also obtained \cite{Airy11}. It has been observed that several dark energy (DE) models with ultra-negative mode equations, as referenced in \cite{Airy12, Airy13, Airy14, Airy15}, provide a better fit with the aforementioned data. These approaches support the Phantom DE scenario, where the constant state parameter equation is utilized, as indicated in \cite{Airy16, Airy17}. The origin of phantom fields is a subject of debate, although they naturally arise in certain models of string theory \cite{Airy18}, supergravity \cite{Airy19}, and theories involving more than 11 dimensions, such as F-theory \cite{Airy20}.\\
Considering that the phantom field is a potential candidate for dark energy, so finding its Black Hole (\textbf{BH}) solutions can be of particular interest, which can be referred to the following references. On the other hand, the regular (singularity free) BHs  are also welcomed by many of people, therefore, the study of  phantom BHs whose singularity is canceled by the phantom field (RPBH) can have a double motivation. In fact, one of the main motivations for embracing regular BHs theories is to avoid space--time structures with incomplete time like geodesics.

In this research, we interest in geodesic paths around the RPBH spacetime, since the geodesics are the free trajectory  of the particles,  providing  new insights into any  spacetime structure.
As mentioned, the RPBH spacetime  structure  are singularity--free, which can be shown to be due to the presence of (phantom) dark energy \cite{Airy21}. Furthermore, Bronnikov and Fabris conducted a study on BHs with self--gravitating phantom scalar fields of arbitrary potentials in vacuum, leading to single out sixteenth classes of possible regular configurations with asymptotically  flat, de Sitter and anti de--Sitter spacetimes\cite{Airy22}.\newline
The study of geodesics holds significant importance, particularly in the exploration of BH spacetime to gain insights into gravitational effects. Investigating the geodesic structure is crucial for verifying new solutions in gravitational theory, presenting a key challenge in astrophysical studies of BHs. Understanding the trajectory of massless/massive particles is essential to  explore the mysteries of BH spacetimes. Researchers have made substantial efforts to explore  the geodesic structure, employing various research methods such as EP analysis method to enhance comprehension in this field \cite{Airy23, Airy24}.\\
In the framework of general relativity, the trajectory of a light ray near a mass distribution is altered due to gravitational forces, causing the light path to curve instead of following a straight line. The degree of light deflection is contingent upon the mass of the gravitational source. Notably, when light traverses near a massive compact object like a BH, the deviation can be significantly pronounced. According to the relativistic viewpoint, gravity curves spacetime, leading to the bending of light as it propagates along the curved spacetime \cite{Airy25}. Recent observational studies conducted by the Event Horizon Telescope have sparked considerable interest among researchers in investigating the optical properties, particularly light bending phenomena, around BHs, including RPBHs \cite{Airy26}.\\
Hence, delving into the study of light deviation around RPBH can provide valuable insights into the spacetime and the structure of the surrounding space. In this work, our focus lies on exploring and analysis of the geodesics of photons and massive particles around RPBH. For better analysis, it is essential to address the concept of EP energy, a powerful tool that helps in comprehending the motion of free particles in the equatorial plane of the spherical  center.  Finally, we examine the light deflection in asymptotically flat RPBH and compare the results  with Schwarzschild BH.\newline
The paper is organized as follows: In Sect.\ref{sec-2} we introduce the regular phantom BH solutions. In Sect. \ref{sec-3} EP are investigated. In Sect.\ref{sec-4},  by geodesic equations, the trajectory of  massive and massless particles are presented. The Sect.\ref{sec-5} is devoted to the  light deflection in asymptotically flat  RPBH.  The conclusions are given in Sect.\ref{sec-6}.
\section{The Regular Phantom Space--Time}\label{sec-2}
\begin{figure}[]
\centering
\includegraphics[width=14cm]{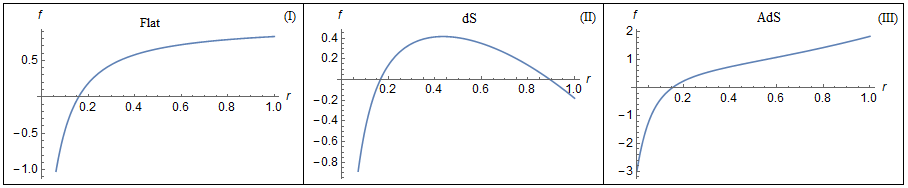}
\caption{The metric coefficient $f(r)$ (\ref{eq3}) for $b=0.01$ and $c=-4$ In the three asymptotically flat, dS, and AdS cases.}
\label{fig1-1}
\end{figure}

\begin{figure}
\centering
\includegraphics[width=14cm]{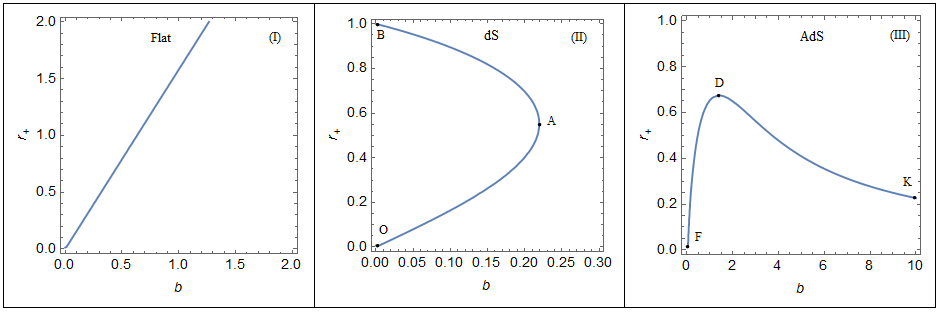}
\caption{The graph of the horizon radius (\ref{eq8}) of the RPBH ($r_{+}$) versus scale parameter ($b$), for $c = -4$. The three sub–figures correspond to the three background spacetimes.}
\label{fig2-2}
\end{figure}

A convenient action, including an arbitrary potential $V(\phi )$  formed by a phantom scalar field  $\phi$, can be written as \cite{Airy7, Airy22}:
\begin{equation}
S=\int{\sqrt{-g}\mathop{dx^{4}}\Big(R+\varepsilon \mathop{g^{\mu \nu }}\mathop{\partial _{\mu }}\phi \mathop{\partial _{\nu }}\phi -2V(\phi )\Big)},
\label{eq1}
\end{equation}
where $R$ is the scalar curvature, $\varepsilon =+1$  describes the usual scalar field with positive kinetic energy and $\varepsilon =-1$ corresponds to the phantom scalar field. To get started, consider a  phantom case ($\varepsilon =-1$) and the following Spherically Symmetric  metric:
\begin{equation}
\mathop{ds^{2}}=f(r)\mathop{dt^{2}}-\frac{\mathop{dr}^{2}}{f(r)}-\mathop{p^{2}}(r)\Big(\mathop{d\theta ^{2}}+\mathop{\sin \theta ^{2}}\mathop{d\varphi ^{2}}\Big),
\label{eq2}
\end{equation}
where  $f(r)$ and $p(r)$ are the unknown metric functions determined by the field equations. By variation of the action (\ref{eq1}) and solving the corresponding field equations,  the metric functions, phantom field $\phi$ and  potential $V$ are obtained as \cite{Airy7, Airy22}:
\begin{equation}
f(r)={p^{2}}(r)\left(\frac{c}{b^{2}}+\frac{1}{p^{2}(r)}+\frac{3M}{b^{3}}\left(\frac{br}{p^{2}(r)}+\arctan \Big(\frac{r}{b}\Big)\right)\right),
\label{eq3}
\end{equation}
\begin{equation}
\mathop{p^{2}}(r)=\mathop{r^{2}}+\mathop{b^{2}},
\label{eq4}
\end{equation}
and
\begin{align}
\phi (r)=\sqrt{2}\in \arctan \Big(\frac{r}{b}\Big)+{{\phi }_{0}},\nonumber\\&V(\phi(r))&=-\frac{c}{{b^{2}}}\frac{{p^{2}}+2{r^{2}}}{{p^{2}}}-\frac{3M}{{b^{3}}}\left(\frac{3br}{{p^{2}}}+\frac{{p^{2}}+2{r^{2}}}{{p^{2}}}\arctan \Big(\frac{r}{b}\Big)\right).
\label{eq5}
\end{align}
The metric function (\ref{eq3}) includes three parameters ($M$, $c$, $b$), the first two are integration constants and the third is a (positive) scale parameter. Regarding these parameters, the following are important:\\
1) $M$ plays the role of the BH mass.\\
2) For $c \geqslant 0$, the metric function (\ref{eq3}) has not any root, means there is no a BH, then in this work, we consider  $c < 0$.\\
3) The parameter $b$  determines the connecting strength between the phantom scalar field and gravity. It is also to note that   as $b\rightarrow0$,  the  Schwarzschild type solution is recovered.\\
4) The asymptotic behavior ($r\rightarrow\infty$) of the metric function (\ref{eq3}) bounds values of the parameters. Three known asymptotically spacetimes considered here are: Flat, dS  and AdS cases. Depending on these cases, the parameters are constrained as follows \cite{Airy22}:\\
for the asymptotically flat case:
\begin{align}
c=-\frac{3\pi M}{2b}
\label{eq6}
\end{align}
and for the asymptotically AdS and dS cases:
\begin{align}
c=\frac{\pm 2{{b}^{3}}-3\pi M}{2b}
\label{eq7}
\end{align}
where the plus (minus) sign is correspond to AdS (dS) case.\\
The graph of  the metric function (\ref{eq3}) versus radial coordinate (in  three asymptotically cases) is illustrated in  Fig.\ref{fig1-1},  for the chosen values $b=0.01$ and $c=-4$.\\
Knowing that  the intersection  point of the graph with the horizontal axis is the horizon radius, we can say that there is always one and only one horizon in the flat  and AdS cases, but  the dS case can have two horizons (as illustrated in Fig.\ref{fig1-1}--middle panel).
The horizon radius $r_+$ can be obtained by vanishing the metric function (\ref{eq3}), that is
\begin{equation}
f(r_+)=1+\frac{c}{b^2}p^2(r_+)+\frac{3Mr_+}{b^2}+\frac{3M}{b^3} p^2(r_+)\arctan \left(\frac{r_+}{b}\right)=0.
\label{eq8}
\end{equation}
On the other hand, choosing an allowed  value for $c$, the mass $M$ becomes a function of $b$ through equations (\ref{eq6}) and (\ref{eq7}).
By taking this in (\ref{eq8}), it turns out that, equation (\ref{eq8})  describes the horizon radius $r_+$ as an implicit function of $b$. The graph of this function is illustrated in Fig.\ref{fig2-2}, for three asymptotically cases as follows:\\
1) Flat case (left panel): the horizon is a linear function in $b$.\\
2) dS case (middle panel): First it is necessary to not that the allowed part of the graph is between two points O and A. This is for the following two reasons:\\
I) The horizon radius is a real  single--valued function.\\
II) As noted above,  RPBH $\rightarrow$  Schwarzschild type BH, as $b\rightarrow 0$, means that dS and flat cases are like as $b\rightarrow 0$.\\
So, the acceptable part of the graph in Fig.\ref{fig2-2} indicates that the scale parameter values are limited to the following range:
\begin{equation}
0<b< b_{max}=b_A\simeq 0.22,
\label{eq9}
\end{equation}
note that the horizon is also restricted between a minimum and a maximum, that is $0<r_+< r_{+max}=r_{+A}$. This means that the dS--RPBH can be formed only for limited values of the scale parameter.\\
3) In AdS case,  the horizon is an increasing function until reaching a maximum (point D), then it decreases monotically to an  asymptotic  value (point K), that is $\lim_{b\rightarrow \infty}r_+=r_{+K}$.\\
Here the physical meaning of the scale parameter may be better understood. Unlike the previous two cases,  the horizon radius (after passing through its maximum) asymptotically approaches a minimum $r_{+K}$, meaning that by increasing  the scale parameter  (after the maximum point (D)), BH becomes smaller.
\section{Effective Potential}\label{sec-3}
EP method is a powerful tool for describing the motion of a free particle in the equatorial plane of a Schwarzschild--like center of attraction. The graph of EP reveals key features of particle motion, serving as a classification scheme for sorting orbits. From the properties of EP, one can derive several interesting qualitative results about the orbits. For instance, stable (unstable) circular orbits in the equatorial plane correspond to the minimum (maximum) points of EP. EP plays a crucial role in geodesic motion, like to the potential in classical mechanics for one--dimensional motion.

In this section, we employ EP method to analyze the motion of photons and massive particles around the RPBH.
For SSS spacetime given by line element (\ref{eq2}), EP is given by \cite{Airy27}:
\begin{equation}
{{V}_{eff}}=\Big(\epsilon+\frac{{{r}^{2}+b^{2}}}{{{r}^{4}}}L^2\Big)f(r),
\label{eq10}
\end{equation}
where $\epsilon=0, 1$ corresponds to the massless and massive particle, respectively and $L$ is the angular momentum constant. Note that the roots of EP  (that is $V_{eff}(r)=0$) are the same BH  horizons.

\begin{figure}[]
\centering
\includegraphics[width=14cm]{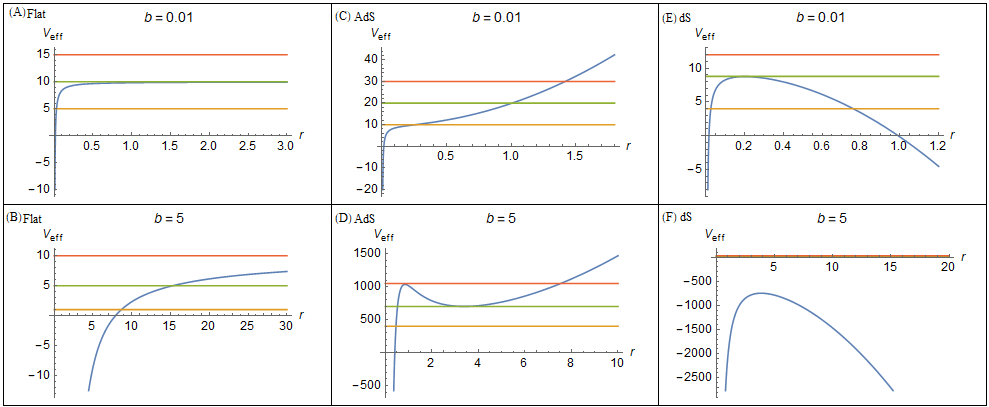}
\caption{The diagram of EP (\ref{eq10}) versus $r$ in three background spacetimes, for $c=-4$.}
\label{fig3-3}
\end{figure}

The graphs of EP (\ref{eq10}) (with $\epsilon=0,  L=3$)\footnote{The graphs for massive particle ($\epsilon=1$) are qualitatively similar.} are plotted in Fig.\ref{fig3-3},  for the metric function (\ref{eq3}). The left, middle and right  panels are correspond to  Flat, AdS and dS cases, respectively. The figure  presents the potential energy diagram for two typical, relatively small  (namely $b=0.01$, upwards  panels) and relatively large (namely $b=5$, downwards panels).
As the figure shows in the Flat case for  relatively small and large  values of scale parameter, the EP is an increasing function from  large negative value to asymptotically  positive value, indicating there is no (relative) extremum point (i. e. no circular orbit). In the AdS case for small values, the situation is
is similar to the Flat case, but, for  the  large values, there is a maximum and a minimum corresponding to unstable and stable circular orbit, respectively. Note that the radius of the stable orbit is greater than the unstable one.

Finally, in the dS case for small values, there is always a maximum point  indicating an unstable circular orbit. For the large  values, the EP is always negative indicating it has  no root  (i. e. no BH).
\section{Geodesics}\label{sec-4}
\begin{figure}[]
  \centering
  \includegraphics[width=14cm]{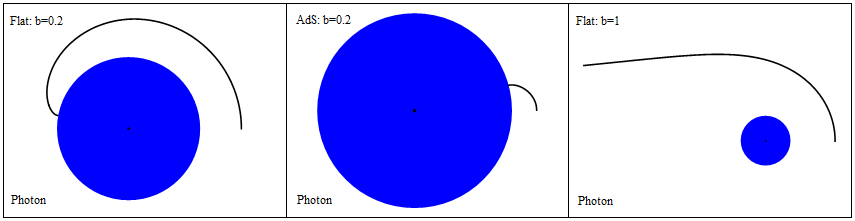}
  \caption{Geodesic paths of photon with the metric coefficient (\ref{eq3}), for  $c=-4$, $L=3$ and different values of $b$. The flat case: $b=0.2$, $r_{h}=0.32$, $r_{0}=0.5$ and $b=1$, $r_{h}=1.6$, $r_{0}=5$.; The AdS case: $b=0.2$, $r_{h}=0.28$, $r_{0}=0.52$. The radius of the blue disk represents the event horizon $r_{h}$.}
  \label{fig4-4}
\end{figure}

\begin{figure}[]
  \centering
  \includegraphics[width=14cm]{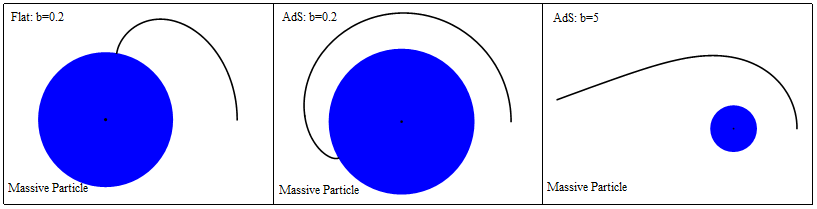}
  \caption{Geodesic paths of massive particle with when the metric coefficient (\ref{eq3}), for $c=-4$, $L=3$ and different values of $b$. The flat case: $b=0.2$, $r_{h}=0.32$, $r_{0}=0.5$. The AdS case: $b=0.2$, $r_{h}=0.28$, $r_{0}=1.6$; and $b=5$, $r_{h}=0.4$, $r_{0}=2$.}
  \label{fig5-5}
\end{figure}

In a general Riemann space, the concepts of straight lines and parallel vectors must be redefined from those familiar in Euclidean geometry. Geodesics serve as generalizations of straight lines in curved spacetime, offering a fundamental tool for a deeper understanding of spacetime and the dynamics of objects within it, thus playing a central role in general relativity. An illustrative example of the significance of geodesics in general relativity is the problem of free falling motion. When the wavelength of gravitational waves is significantly smaller than the radius of curvature of the background spacetime, these waves propagate along the geodesics of the metric. Geodesy aims to determine the path of free particle motion in the gravitational field of a given spacetime, or more precisely, to ascertain the structure of the gravitational  field. Additionally, from an astronomical perspective, geodesics are crucial tools for studying phenomena such as light deflection by galaxies acting as massive gravitational lenses.\\
Let's begin with the  full geodesic equation given  by:
\begin{equation}
\mathop{{\ddot{x}}^{\alpha }}+\mathop{\Gamma _{\beta \nu }^{\alpha }}\mathop{{\dot{x}}^{\beta }}\mathop{{\dot{x}}^{\nu }}=0,
\label{eq11}
\end{equation}
where $\mathop{\Gamma }_{\beta \nu }^{\alpha }$ are affine connections and dot denotes the derivative with respect to an affine parameter. We want to write (\ref{eq11}) for the metric (\ref{eq2}) (specified by the metric function (\ref{eq3})). In the SSS metric case, without loss of generality, we can study the motion  in equatorial plane ($\theta =\frac{\pi }{2}$). Imposing this condition, the motion  equations (\ref{eq11}), for the metric (\ref{eq2}), take the following form:
\begin{equation}
\ddot{t}+\Big(\frac{f'(r)}{f(r)}\Big)\dot{t}\dot{r}=0,
\label{eq12}
\end{equation}
\begin{equation}
\ddot{r}+\frac{1}{2}f(r)f'(r)\mathop{{\dot{t}}^{2}}-\frac{1}{2}\frac{f'(r)}{f(r)}\mathop{{\dot{r}}^{2}}-f(r)p(r)p'(r)\mathop{{\dot{\phi }}^{2}}=0,
\label{eq13}
\end{equation}
\begin{equation}
\ddot{\phi}+\frac{2 p'(r)}{p(r)}\dot{r}\dot{\phi}=0,
\label{eq14}
\end{equation}
where prime denotes the derivative with respect to $r$. The above equations can not be solved exactly  and so  we must  use the numerical methods. To do this, the initial conditions are also required which are fixed as
\begin{center}$r(0)=r_{0}$\rlap.\footnote{$r_{0}$ is the initial location that is placed at a suitable distance from the event horizon ($r_{0}>r_{h}$).} , $\phi(0)= 0$, $\dot{r}(0)=0$ and $L=3$.\end{center}
After implementing the numerical commands, the outputs for the geodesic paths  appear as shown in Fig.\ref{fig4-4} and Fig.\ref{fig5-5}, for photon and massive particles respectively. More detailed descriptions for Fig.\ref{fig4-4} and Fig.\ref{fig5-5} are as follows:\newline
1) The comparison between the geodesics of photon and massive particle  in the flat case (left panels in Figs \ref{fig4-4}, \ref{fig5-5}) shows that , under similar conditions the massive particle travels a shorter path than a photon before falling into a BH. This means that gravity is stronger on massive particles.\\
2) The comparison between the geodesics of   photon and massive particle  in the AdS case (middle panels in Figs \ref{fig4-4}, \ref{fig5-5}) shows that , under similar conditions the photon travels a shorter path than massive  before falling into a BH. This means that gravity is stronger on photon (exact  opposite of the Flat case).\\
3) The right panels in Figs \ref{fig4-4}, \ref{fig5-5} are examples of  conditions where photon ans massive particle are able to escape the gravity field of the BH.\\
At the end of this part, we will point out two points. First, as the $b$  parameter increases, the paths in the left and middle panels of Figs \ref{fig4-4}, \ref{fig5-5} become shorter, indicating the stronger gravity. Second, the software calculations for plotting the graphs in the dS case have no output.\\
\begin{center}\textbf{Circular Orbits}\end{center}
Circular motion are the simplest form of orbital motion in celestial mechanics, characterized by a rotating body maintaining a constant radius while orbiting a gravitational mass. These orbits are especially significant in the equatorial plane, such as in the structure of accretion disks around astrophysical BHs. EP is a valuable analytical tool for describing the motion of a free particle within this plane. It encapsulates many fundamental aspects of particle dynamics, serving as a classification scheme for orbit configurations. Circular orbits  are possible if  $\frac{dV_{eff}}{dr} = 0$, which determines the minimum and maximum points of EP. The stable (unstable) circular orbits are corresponds to minimum (maximum)  points.

By examining Fig.\ref{fig3-3}, we can see that in  the flat case,  there is no extremum point, meaning that the formation  of  circular orbits  is ruled out. However, in AdS case (for values greater than a certain value of the scale parameter), there exist one stable (corresponding to minimum) and one unstable (corresponding to maximum) orbit. Finally, in dS case,  there is  always only  an  unstable orbit.

A remarkable feature  of  circular orbits is the orbital frequency.  It is  one of  the fundamental quantities  in the motion such that  it can be used to analyze or obtain many properties of motion or objects involved in motion. One example of astrophysical cases is the determination of invisible objects in binary systems. Therefore, we will focus our attention on it in the following.

\begin{figure}[H]
  \centering
  \includegraphics[width=14cm]{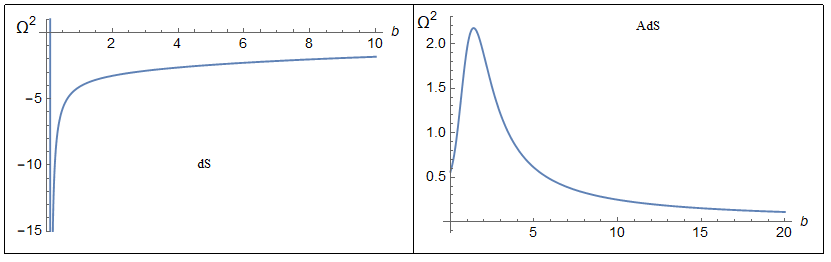}
  \caption{Orbital frequency $\Omega$ (\ref{eq20}) versus $b$ in the three sub–figures correspond to the three background spacetimes, for numerical values of parameters $c=-4$ and $L=3$.}
  \label{fig6-6}
\end{figure}

To calculate the  orbital frequency, we  begin with the equation that determines the potential extremum, that is
$\frac{dV_{eff}}{dr} = 0$, which by substituting from equation (\ref{eq10}), one gets:
\begin{equation}
\frac{d{{V}_{eff}}}{dr}=f'(r)(\varepsilon +{{L}^{2}}(\frac{{{r}^{2}}+{{b}^{2}}}{{{r}^{4}}}))-f(r)(2{{L}^{2}}(\frac{{{r}^{2}}+2{{b}^{2}}}{{{r}^{5}}}))=0,
\label{eq15}
\end{equation}
after simplification, we have
\begin{equation}
\frac{f'(r)}{f(r)}=\frac{2{{L}^{2}}({{r}^{2}}+2{{b}^{2}})}{r(\varepsilon r^{4} +{{L}^{2}}({{r}^{2}}+{{b}^{2}}))}.
\label{eq16}
\end{equation}
Also, for circular motion, by putting $\dot{r}=\ddot{r}=0$, $\dot{\phi}=\omega$ in  equation (\ref{eq14}), we get
\begin{equation}
\frac{1}{2}f(r)f'(r){{\dot{t}}^{2}}-f(r)p(r)p'(r){{\omega }^{2}}=0,
\label{eq17}
\end{equation}
on the other hand, from equation (\ref{eq13}), it is easily to obtain $\dot{t}=\frac{E}{f(r)}$ which by substituting in the last equation, we obtain
\begin{equation}
{{\omega }^{2}}={{E}^{2}}\frac{f'}{2{{f}^{2}}pp'},
\label{eq18}
\end{equation}
finally by eliminating the derivative of the metric function ($f'$) from equations (\ref{eq16}) and (\ref{eq19}), one obtain the main important equation:
\begin{equation}
{{\omega}^{2}}={{L}^{2}}{{E}^{2}}\frac{({{r}^{2}}+2{{b}^{2}})}{rp(r)p'(r)(\varepsilon {r^{4}}+{{L}^{2}}({{r}^{2}}+{{b}^{2}}))f(r)}={{L}^{2}}{{E}^{2}}\frac{({{r}^{2}}+2{{b}^{2}})}{2r^2(r^2+b^2)(\varepsilon {r^{4}}+{{L}^{2}}({{r}^{2}}+{{b}^{2}}))f(r)},
\label{eq19}
\end{equation}
or in terms of new (rescaled) orbital frequency  $\Omega= \omega/LE$:
\begin{equation}
{{\Omega}^{2}}=\frac{{{r}^{2}}+2{{b}^{2}}}{2r^2(r^2+b^2)(\varepsilon {r^{4}}+{{L}^{2}}({{r}^{2}}+{{b}^{2}}))f(r)}.
\label{eq20}
\end{equation}
Note that in the last equation, the quantity $r$ represents the orbital radius which is an implicit function of  $b$ given by (\ref{eq15}), therefore, the rescaled orbital frequency (\ref{eq20}) becomes a function of $b$. By substituting $f(r)$ from equation (\ref{eq3}), we can plot this function illustrated in  Fig.\ref{fig6-6} for the dS (left panel) and AdS (right panel) cases.

As the figure shows the dS case represents an unphysical situation, since for a wide range of scale parameters, the orbital frequency is negative.
This means that the (unstable) circular orbits are confined to  a very small range  of scale parameter. And even for that very small range, the orbital frequency changes are extremely sharp which isn't physically acceptable.

In AdS case, the orbital  frequency is an increasing function of the scale parameter $b$ until it reaches a maximum, after which the frequency decreases relatively rapidly despite the increase in the parameter.  This physically means that increasing the parameter to the maximum point makes gravity stronger, and after that, despite the increase in the parameter, gravity becomes weaker. This behavior is qualitatively similar to the behavior of the horizon radius in Fig.\ref{fig2-2}. Recall that the scale parameter determines the  connecting strength between the phantom scalar field and gravity.
\section{Light Deflection}\label{sec-5}
\begin{figure}[H]
\centering
\includegraphics[width=14cm]{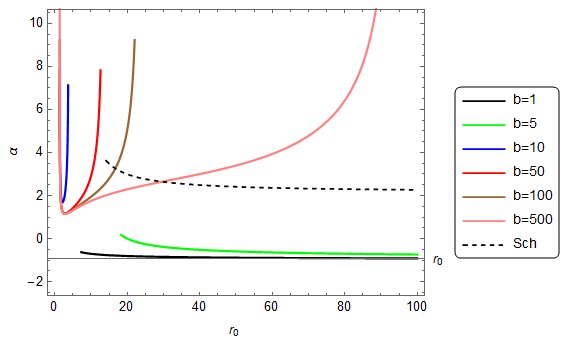}
\caption{The light deviation (\ref{eq28}) of RPBH in asymptotically flat case for different value of $b$ and $c=-4$.  The corresponding  Schwarzschild graph is also shown for comparison.}
\label{fig7-7}
\end{figure}

The idea that light can be deflected due to gravity has ancient origins. It has become a cornerstone of the empirical evidence that supports general relativity. Indeed, it plays a main role in Gravitational Lensing topic which  has become a major tool in efforts to understand how matter is distributed around galaxies and clusters, and in searches for DM, DE, compact objects, and extrasolar planets. Historically, \emph{Henry Cavendish}, proposed that the mass of a star could be estimated by measuring how the speed of light would slow down due to the star's own gravitational field.
Classically, it has been shown that a ray of light, which passes at a known distance from a mass, undergoes a deflection. As if this would have been possible for the sun.
Accordingly, study of light deflection around the massive object such as BH is important because of extracting information such as the curvature of spacetime and the mass distribution.
A lot of valuable work done in this field is specific to BHs, both static and rotating. Specially, in the case of SSS and asymptotically flat spacetime an analytic formula is obtained. In this work, we use this formula for calculation of the light  deflection for   asymptotically flat RPBH. It is shown that for the following  SSS:
\begin{equation}
ds^2=B(r)dt^2- A(r)dr^2-D(r)r^2\Big(d\theta^2+\sin^2\theta d\phi^2\Big),
\label{eq21}
\end{equation}
the deflection angle is given by \cite{Airy28}:
\begin{equation}
\alpha =2\mathop{\int }_{r_0}^{\infty }\sqrt{\frac{A\left( r \right)}{D\left( r \right)}}\sqrt{{{\left( \frac{r}{{{r}_{0}}} \right)}^{2}}\frac{D\left( r \right)}{D\left( {{r}_{0}} \right)}\frac{B\left( {{r}_{0}} \right)}{B\left( r \right)}-1}\frac{dr}{r}-\pi
\label{eq22}
\end{equation}
where $r_{0}$ is the distance of closest approach of the light to the center of the gravitational attraction.
The Integral Formula is done by Weinberg. Solving the complex metrics is difficult by this formula, therefor Amore et al \cite{Airy29} by the mathematic methods change this integral formula to a more convenient formula.\\
Amore et al performed a change of variable, $z=\frac{r_{0}}{r}$, and introduced a function:
\begin{equation}
V(z)={{z}^{2}}\frac{D(\frac{{{r}_{0}}}{z})}{A(\frac{{{r}_{0}}}{z})}-\frac{{{D}^{2}}(\frac{{{r}_{0}}}{z})B({{r}_{0}})}{A(\frac{{{r}_{0}}}{z})B(\frac{{{r}_{0}}}{z})D({{r}_{0}})}+\frac{B({{r}_{0}})}{D({{r}_{0}})},
\label{eq23}
\end{equation}
if the coefficients $A$, $B$ and $D$ are replaced in (\ref{eq23}), it will be converted as follows:
\begin{equation}
V(z)=\sum\limits_{n=1}^{\infty }{{{v}_{n}}{{z}^{n}}}={v_{1}}{z^{1}}+{v_{2}}{z^{2}}+{v_{3}}{z^{3}}+... .
\label{eq24}
\end{equation}
On the other hand, they defined a ”transformed” potential:
\begin{equation}
\rho (z)=\sum\limits_{n=1}^{\infty }{{{v}_{n}}{{\Omega }_{n}}{{z}^{n}}},
\label{eq25}
\end{equation}
where
\begin{equation}
{{\Omega }_{n}}=\sqrt{\pi }\frac{\Gamma (\frac{n}{2}+\frac{1}{2})}{\Gamma (\frac{n}{2})}.
\label{eq26}
\end{equation}
By performing mathematical operations and simplifying Eq.(\ref{eq22}), it becomes as follows:
\begin{equation}
\alpha =\pi \Big[\sqrt{\frac{\pi }{2\rho (z)}}-1\Big]
\label{eq27}
\end{equation}
For $z=1$, $\rho (1)$ is obtained by Eq.(\ref{eq25}).\\
By substituting $\rho (1)$ into Eq.(\ref{eq27}), the deflection angle of RPBH is equal to:
\begin{equation}
\alpha =-1+\sqrt{2}{{\pi }^{3/2}}\sqrt{\frac{X}{Y}},
\label{eq28}
\end{equation}
where  $X$ and $Y$ are:
\begin{equation}
X=-{{b}^{2}}{r_{0}^{2}}\Big( {{b}^{2}}+{r_{0}^{2}} \Big),
\label{eq29}
\end{equation}
\begin{equation}
\resizebox{0.9\textwidth}{!}{%
$Y=-32M{r_{0}^{5}}+2{{b}^{2}}{r_{0}^{3}}\Big( -40M+( -1+6M)\pi r_{0} \Big)+{{b}^{4}}\Big( 9M\pi -48Mr_{0}-\pi {r_{0}^{2}} \Big)+3bM\pi \Big( {{b}^{2}}+{r_{0}^{2}} \Big)\Big( 3{{b}^{2}}+4{r_{0}^{2}} \Big)\arctan \Big( \frac{r_{0}}{b} \Big)$}.
\label{eq30}
\end{equation}
The variations of angel deviation $\alpha$ versus $r_0$  are plotted in Fig.\ref{fig7-7} for different values of $b$. As the figure shows, approximately for
$r_0 > 5$, the deviation angle can be an asymptotically (negative) decreasing function of slow variations (for typical value $b=5$), or it can be   an positive increasing function  (for typical value $b=500$). Note that for small values of $b$, the behavior is Schwarzschild--like.
\newpage

\section{Conclusion}\label{sec-6}
The RPBH is described by the action involving  a self--gravitating scalar field and arbitrary potential.
In this manuscript, some important geodesic properties of  RPBH for three asymptotically cases, flat, dS and AdS  are examined.

In the meantime, the effects of the important scale parameter (which determines the coupling strength  between gravity and scalar field) on the geodesic paths  for massless and massive particles are presented. The most important results can be stated as follows:\\
1) In the flat case and  under similar conditions, massive particles falling into a BH take a shorter path than massless particles.\\
2) In the AdS case and  under similar conditions, massless  particles falling into a BH take a shorter path than massive particles.\\
3) In  the dS  case, as far as the authors' software calculations are concerned, the display of paths has no output.\\
4) Using the EP method, it is shown that the formation of circular orbits in the Flat case is impossible and in the dS case leads to unphysical results. But in the AdS case, circular orbits, beyond a certain value for the scale parameter, stable and unstable circular orbits can be formed. The graph of the  angular frequency of the circular orbits shows that, it  becomes maximum at a certain value the scale parameter, e. g. $ b_{max}$ and then decreases. In other words, in the interval $0<b< b_{max}$, the angular frequency is  an increasing function of $b$,  and then decreasing. This physically means that in the mentioned interval, the  gravity becomes stronger  and then begins to weaken (relatively quickly), despite the $b$ increases.

\end{document}